\documentclass[english,prb,aps,article,twocolumn,showpacs,reprint,microtype,longbibliography,superscriptaddress]{revtex4-1}

\usepackage{graphicx}
\usepackage{subfig} 
\usepackage{dcolumn}
\usepackage{multirow}
\usepackage{threeparttable} 
\usepackage{soul} 
\usepackage{verbatim} 

\def\beq{\nopagebreak \begin{equation}}
\def\eeq{\end{equation}}

\usepackage{amsmath}
\usepackage{gensymb}
\usepackage{textgreek}
\usepackage{caption}
\usepackage{braket}
\usepackage{color}
\definecolor{forestgreen}{RGB}{34, 139, 34}
\usepackage[unicode=true,pdfusetitle,
 bookmarks=true,bookmarksnumbered=false,bookmarksopen=false,
 breaklinks=false,pdfborder={0 0 1},backref=false,colorlinks=true]
 {hyperref}
 \usepackage{siunitx}

\begin{document}
\title{Thermal resistance of GaN/AlN graded interfaces}

\author{Ambroise van Roekeghem}
\email{ambroise.vanroekeghem@cea.fr}
\affiliation{CEA, LITEN, 17 Rue des Martyrs, 38054 Grenoble, France}

\author{Bjorn Vermeersch}
\altaffiliation[Current address: ]{imec, Kapeldreef 75, 3001 Leuven, Belgium}
\affiliation{CEA, LITEN, 17 Rue des Martyrs, 38054 Grenoble, France}

\author{Jes\'{u}s Carrete}

\affiliation{Institute of Materials Chemistry, TU Wien, A-1060 Vienna, Austria}

\author{Natalio Mingo}

\affiliation{CEA, LITEN, 17 Rue des Martyrs, 38054 Grenoble, France}

\date{\today}
\begin{abstract}
Compositionally graded interfaces in power electronic devices eliminate dislocations, but they can also decrease thermal conduction, leading to overheating. We quantify the thermal resistances of GaN/AlN graded interfaces of varying thickness using {\it ab initio} Green's functions, and compare them with the abrupt interface case. A non-trivial power dependence of the thermal resistance versus interface thickness emerges from the interplay of alloy and mismatch scattering mechanisms. We show that the overall behavior of such graded interfaces is very similar to that of a thin-film of an effective alloy in the length scales relevant to real interfaces.
\end{abstract}

\maketitle

\section{Introduction}

The common strategy of interfacing two pure semiconductors via a gradual alloyed region poses a design trade-off: on the one hand, a thicker alloyed region is desirable in order to accommodate the lattice mismatch; on the other hand, a thick alloy will also negatively impact the thermal conductance across the interface, possibly degrading a device's lifetime. For example, common substrate designs for High Electron Mobility Transistors (HEMTs) include a graded transition region between GaN and AlN as thick as \SI{1.5}{\micro m} \footnote{C. Giesen, private communication}. For substrates grown on Si, this region contributes only 3\% of the total substrate thermal resistance. But if one replaces Si by diamond, the graded interface thermal resistance amounts to 20\% of the total. For common operating conditions in HEMTs this translates into a \SIrange[range-phrase=--, range-units=single]{10}{20}{\celsius} higher temperature of the active region, shortening the device's lifetime by one half \footnote{Assuming \SI{1000}{\micro m} of Si or diamond, and \SI{7}{W/(m.K)} for the graded region}.

Thus, it may pay off to try to make the graded region as thin as possible while still satisfying some minimum structural constraints. To do this one needs to know how the interface thermal resistance depends on its thickness and composition profile. Theoretically quantifying it is, however, a nontrivial problem due to the simultaneous emergence of quasiballistic transport, wave interference effects, and the interplay between large-scale features and atomic-scale disorder. All this leads to a thermal resistance that no longer depends linearly on thickness when the latter gets below the \si{\micro m} range. Experimentally one can only access the thermal conduction of the structure as a whole, which typically includes many different material interfaces in a single sample. This makes it difficult to quantify the upper limits to interface conductance based on experiments alone. It is therefore very important to theoretically predict the thermal conductance of graded interfaces, to know how much the grading strategy is degrading the performance of devices, and how much improvement could be achieved by other schemes, like reducing thickness or using digital structures. We tackle this question here for the technologically important case of GaN/AlN interfaces, which are ubiquitous in power electronics where thermal dissipation issues are a major concern.

The use of graded interfaces to reduce threading dislocation density was theoretically tackled by Tersoff using an analytical model, and has been deeply explored in subsequent modeling studies for two decades \cite{tersoff_dislocations_1993,bertoli_misfit_2009}. The specific grading length and profile to satisfy structural constraints (e.g. dislocation density or strain) depends on the particular problem in hand. The structural constraint aspect is not the goal of this paper, and we refer the reader to earlier literature. 

To model the thermal conductance of interfaces, mainstream techniques include the use of different phonon-gas models such as the diffuse mismatch or acoustic mismatch models, atomistic Green's functions, or molecular dynamics\cite{Swartz_Pohl_review,Liang_thin-film_GaN,Gaskins_ZnO_GaN_interface,Ong_tutorial_AGF}. A few theoretical studies have focused on trying to improve the interface thermal conductance between two different materials by using mass gradients\cite{Zhou_mass_graded_interface}. However, in practice, the intermediate layers are often not perfectly ordered in the plane perpendicular to the interface, but rather disordered alloys at a given composition. This is the case that we study here, using Monte Carlo simulations to account for the anharmonic and alloy scattering effects coupled to atomistic Green's functions to compute the probability of transmission at the interface of two materials. We focus on the case of a very progressive step graded interface and show that the total interface can be modeled as a thin film of an effective alloy, due to the dominant contribution from alloy scattering at realistic length scales. According to recent experimental results \cite{Gaskins_ZnO_GaN_interface}, the interface thermal resistance can actually be overestimated by atomistic Green's functions due to additional conductance from non-elastic processes, so this statement should be very robust in the case of those progressive interfaces of two similar crystal structures. We give a general expression for the characteristics of this effective alloy, such that the result can be easily applied to other interfaces profiles.

\section{Methods}
Current approaches to investigate interface thermal conductance atomistically are based on Green's functions \cite{mingo_phonon_2003,mingo_greens_2009,ong_efficient_2015,tian_greens_2014,wang_quantum_2008,zhang_simulation_2007,zhang_atomistic_2010}. However, when the interface is structurally complex or disordered in an extended region of space, like in the case of graded interfaces, it is impractical to simulate the whole atomically disordered interface. Doing so would require approximating the disordered interface by multiple realizations of exceedingly large supercells, periodic in the directions parallel to the interface \cite{tian_greens_2014}. 
A method to overcome this limitation was demonstrated in Refs.~\onlinecite{chen_role_2013,carrete_predictive_2018}. The basic idea is that the  perturbation part of the Hamiltonian can be written as the sum of a contribution from the average perturbation over the layer (the ``compositional profile'' contribution), plus the local deviation with respect to this average for each atom (the ``disorder'' contribution). The total phonon scattering intensity is the sum of the scattering intensities of each of the two contributions separately, plus the interference term between them, which is neglected on the grounds of phase cancellation due to disorder. The ``compositional profile'' scattering term preserves the phonon quasimomentum component parallel to the interface, and can be efficiently calculated using atomistic Green's functions. The ``disorder'' term does not preserve quasimomentum, and it is computed within the framework of the virtual crystal approximation. Using realistic compositional profiles and {\it ab initio} calculated force constants, a good agreement with experiment was obtained for superlattices with periods comprising up to \num{45} atomic layers, without using any adjustable parameters \cite{chen_role_2013,carrete_predictive_2018}.

The approach implemented for superlattices is not directly applicable to single interfaces, and requires modification. The difference comes from the periodic character of the superlattices, versus the local and asymmetric character of the single interface. The repeated structure of the superlattices justifies a picture where phonons travel in an effective medium and are incoherently scattered by the homogeneously distributed changes in composition. The T-matrix formalism is best suited to this problem. In contrast, the materials at each side of the single interface are dissimilar, and thus cannot be be treated by the T-matrix formulation of Refs.~\cite{chen_role_2013,mingo_ab_2014,protik_ab_2016}, which requires having the same material infinitely far from the interface on either side. Instead, the Green's function formalism for transmission and reflection probabilities is used here \cite{mingo_phonon_2003,mingo_greens_2009}.

\begin{figure}
\begin{centering}
\includegraphics[width=8.5cm]{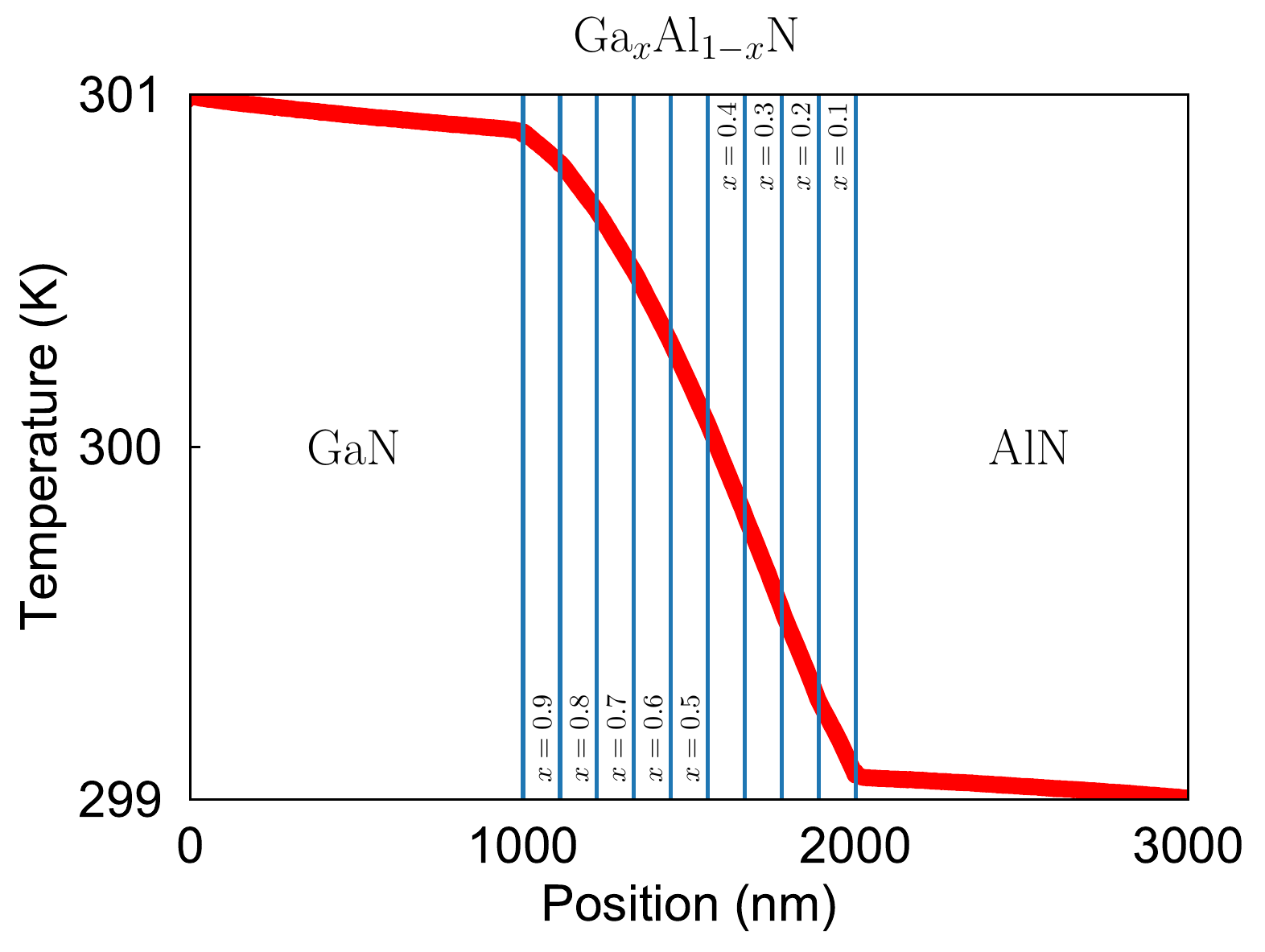}
\par\end{centering}
\caption{Structure of the considered GaN/AlN (001) step graded interfaces and associated temperature profile (here the total length of the interface is \SI{999}{nm}).\label{fig:Temperature-profile}}

\end{figure}

In the considered geometry, the interface is bidimensional and infinite
in two directions, so that a transmission or a reflection event
will conserve the component of momentum parallel to the interface
as well as the energy. The dynamical matrix of each bulk material,
calculated {\it ab initio}, is first Fourier-transformed in the direction
perpendicular to the interface to obtain a set of mixed-space force
constants:
\begin{widetext}
\begin{equation}
\tilde{\Phi}_{\alpha i}^{\alpha'i'}\left(q_{\parallel},X_{m}\right)=\sum_{n<N}\sqrt{M_{\alpha}M_{\alpha'}}D_{\alpha i}^{\alpha'i'}\left(\mathbf{q}_{\parallel}+\mathbf{q}_{\perp_{n} }\right)e^{-iq_{\perp_{n}}X_{m}},
\end{equation}
\end{widetext}

\noindent where $N$ is the chosen number of samples over the Brillouin zone in
the direction perpendicular to the interface, $\alpha$ and $\alpha'$
the indexes over atoms, $i$ and $i'$ over Cartesian directions,
$D$ the dynamical matrix and $M$ the masses of the atomic species.
$X_{m<N}$ are lattice vectors perpendicular to the interface and
$\tilde{\Phi}_{\alpha i}^{\alpha'i'}\left(\mathbf{q}_{\parallel},X_{m}\right)$ corresponds
to the interactions between layers separated by $X_{m}$ in real space
for a given parallel wavevector.

Using this set of force constants, the 2-D Green's function $g\left(\mathbf{q}_{\parallel}\right)$
of each material is computed using the decimation technique \cite{guinea_effective_1983,sancho_highly_1985}. The interactions $k_{12}\left(\mathbf{q}_{\parallel}\right)$ between layers of different materials 1 and 2
in real space are obtained from the averaged mixed-space force constants
and using the masses of the two different materials. For $\mathbf{q}_{\parallel}=0$,
the acoustic sum rule is enforced by correcting the values of the
diagonal elements. Finally, we obtain the transmission between the two materials using the two-region formula\cite{mingo_greens_2009,mingo_phonon_2003}, for each parallel wavevector:

\begin{equation}
T=4\pi^{2}Tr\left[\rho_{1}D_{1}^{A}k_{12}\rho_{2}D_{2}^{R}k_{21}\right]
\end{equation}

With $D_{1}^{A}=\left(I-k_{12}g_{2}^{+}k_{21}g_{1}^{+}\right)^{-1}$, $D_{2}^{R}=\left(I-k_{21}g_{1}k_{12}g_{2}\right)^{-1}$, and $2\pi\rho=g-g^{+}$ the spectral density of states of the decoupled system.

This transmission is subsequently used within the Monte
Carlo solver implemented in almaBTE \cite{carrete_almabte_2017}. The {\it{ab initio}} data for GaN and AlN has been computed in a 5x5x5 supercell within the Local Density Approximation of Density Functional Theory, and can be retrieved from the database of the almaBTE website. For each incoming
mode, only outgoing modes which conserve the parallel wavevector and
the frequency are considered (for transmission or reflection). When several outgoing modes are available for one incoming mode, their probability is weighted by their group velocity projected on the normal to the interface. 

One difficulty in simulating a step graded interface is to combine
this mismatch scattering with the intrinsic scattering from the alloy
disorder within the layers. To tackle this issue, we choose to slice
the interface into steps with gradual increases in composition of 10\%
(see Fig.\ \ref{fig:Temperature-profile}). In the limit of very short interfaces,
mismatch scattering becomes dominant, while if the interface is very
thick, alloy scattering takes over.

Finally, the interface thermal resistance is evaluated from the temperature
profiles and heat fluxes obtained from the Monte Carlo simulation using
buffer layers of \SI{1}{\micro m} at each side of the interface, as:
\begin{equation}
R=\frac{\Delta T}{J},
\end{equation}

\noindent with $J$ representing the total heat flux across the structure and $\Delta T$
the jump in temperature across the interface.

\section{Thermal resistances of GaN/AlN step graded interfaces}

In the following, we focus on step graded interfaces from GaN to AlN, oriented along the (001) direction. All properties are computed {\it ab initio}, using a $15\times 15\times 15$ grid for the phonon wavevectors and $15$ samples for the mixed-space force constants.

\begin{figure}
\begin{centering}
\includegraphics[width=8.0cm]{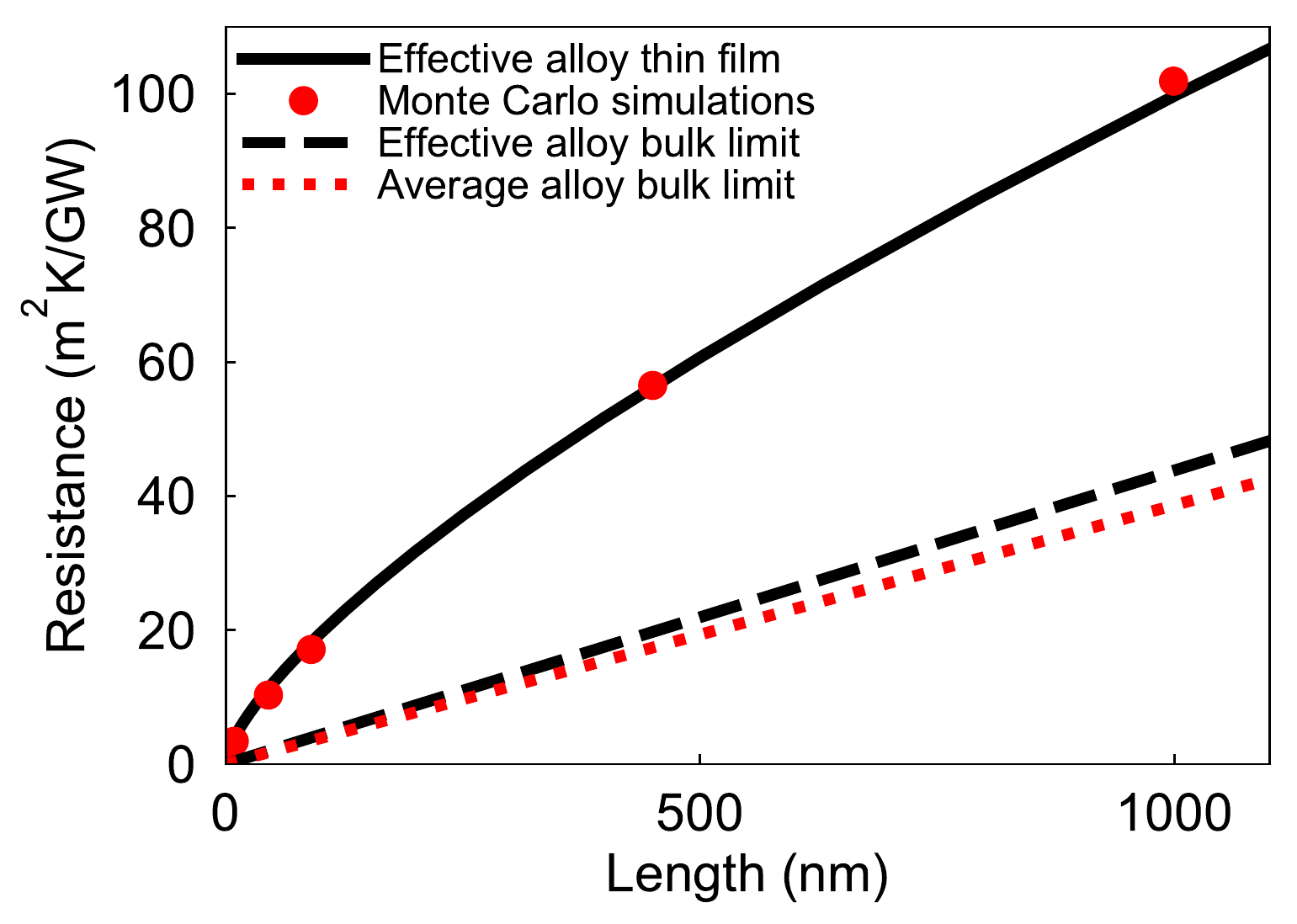}
\includegraphics[width=8.5cm]{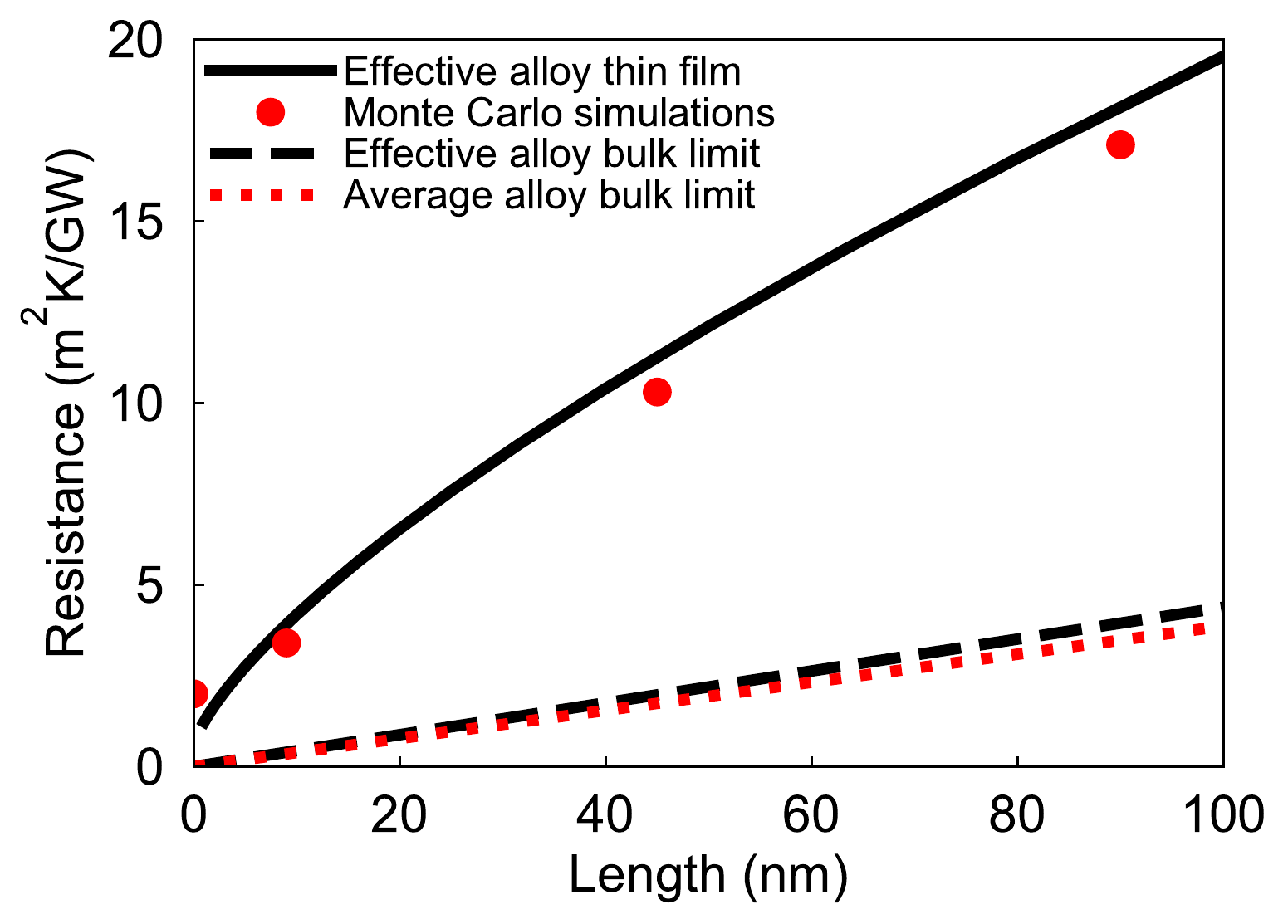}
\includegraphics[width=8.5cm]{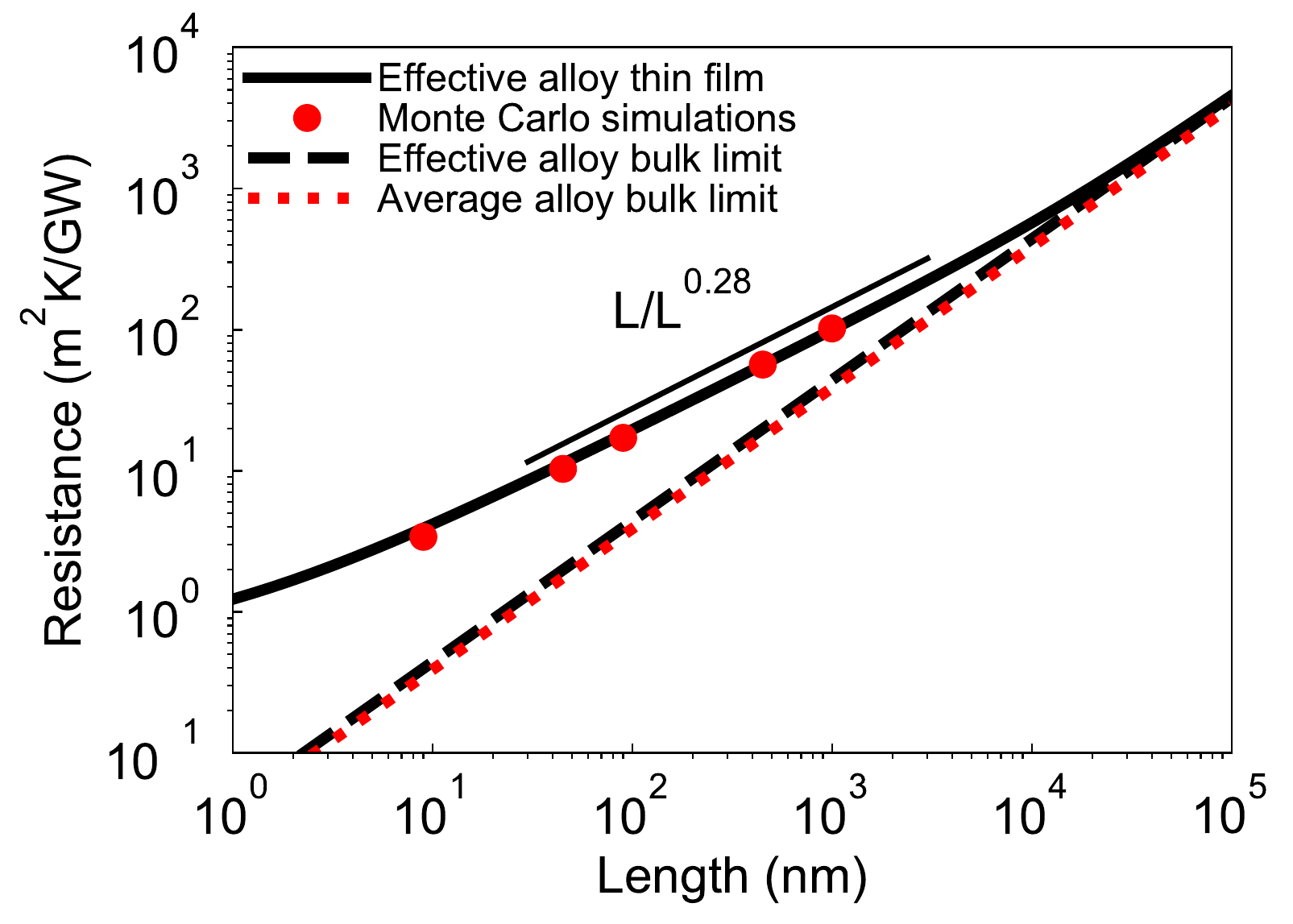}
\par\end{centering}
\caption{Resistances of GaN/AlN (001) step graded interfaces of different lengths as computed from Monte Carlo simulations, compared with the resistance of the effective alloy thin film and with the bulk effective alloy and average alloy limits. Upper panel: Full data, middle panel: zoom at short lengths, lower panel: data in log-log scale to demonstrate the effective power law for the interface resistance.\label{fig:Resistance}}

\end{figure}

In Figure~\ref{fig:Temperature-profile}, we display the temperature profile obtained for a step graded interface of about one micron. As can be seen from the absence of temperature jumps, the effect of the successive interfaces on the thermal resistance is negligible compared to the scattering from alloy disorder. In this context, one can model the step graded structure as an effective alloy in which the mass disorder scattering contribution is similar to that of an alloy with concentration Ga$_{\tilde{x}}$Al$_{1-\tilde{x}}$N such that 

\begin{equation}
\tilde{x}(1-\tilde{x})=\frac{1}{N}\sum_{i<N}x_{i}\left(1-x_{i}\right),
\end{equation}

\noindent where $x_{i}$ the concentration in each layer of the alloy. A similar approach has been shown to give excellent results in the context of superlattices \cite{carrete_predictive_2018}. Here, $\tilde{x}\simeq 0.24$ and in the continuous limit $\tilde{x}=1/2-\sqrt{3}/6$. At the same time, the phonon spectrum, average masses and three-phonon properties of this effective alloy correspond to those of the average alloy (here with concentration Ga$_{0.5}$Al$_{0.5}$N). Finally, the finite length of the interface limits the mean-free path of the phonons, as described in Ref.~\onlinecite{Vermeersch_cross-plane_conduction}:

\begin{equation}
\frac{1}{R} = \sum_{v_{\perp}(q)>0}\frac{C(q)v_{\perp}(q)\Lambda_{\perp}(q)}{L+2\Lambda_{\perp}(q)}
\end{equation}

With $R$ the thermal resistance of the interface of length $L$, $C$ the specific heat of a phonon mode, $v_{\perp}$ and $\Lambda_{\perp}$ its group velocity and mean free path projected on the normal to the interface.
As shown in Figure~\ref{fig:Resistance}, such an approximation gives excellent results in the range of lengths applicable to real interfaces, from nanometer to micrometer scales. We also observe in the lower panel a power law similar to the one found for AlGaN alloys thin films \cite{Vermeersch_cross-plane_conduction}.

\begin{figure}
\begin{centering}
\includegraphics[width=9cm]{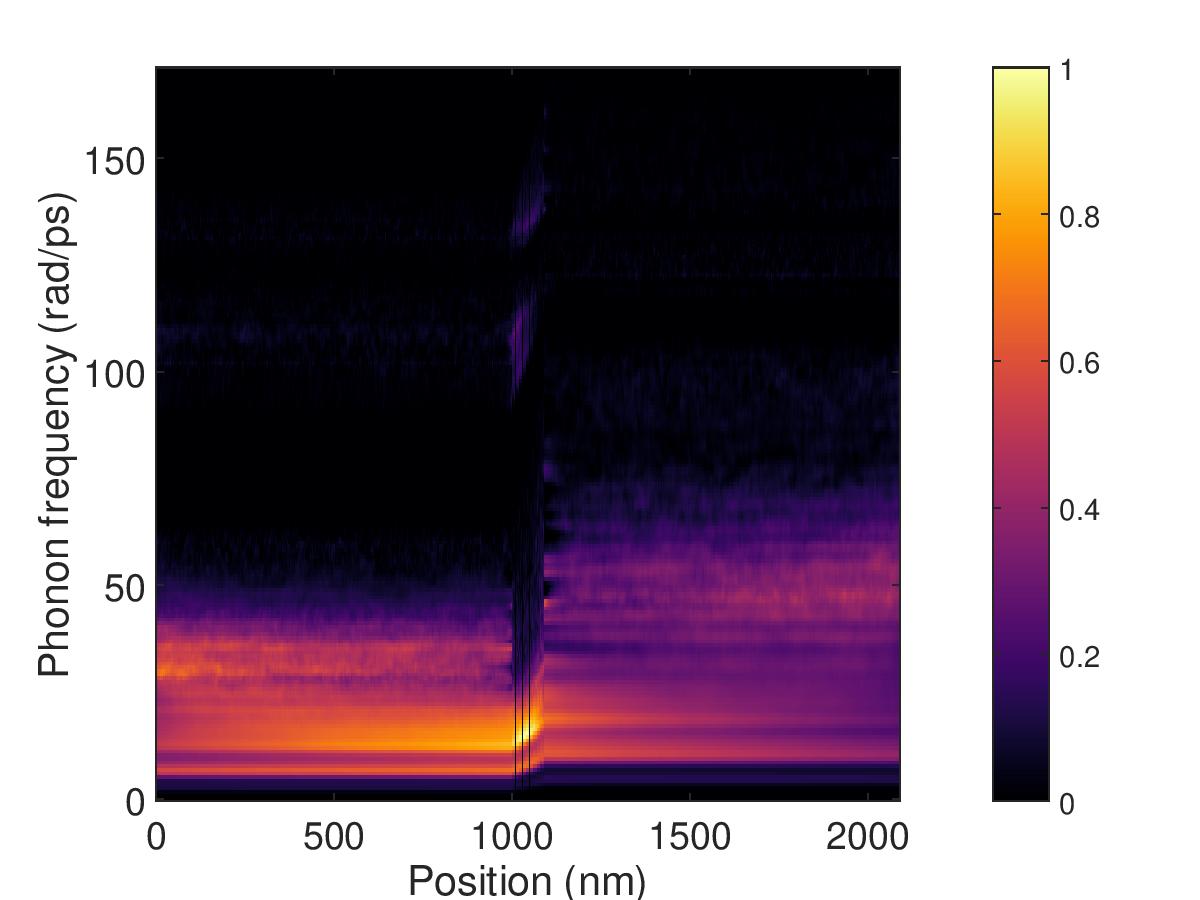}
\includegraphics[width=9cm]{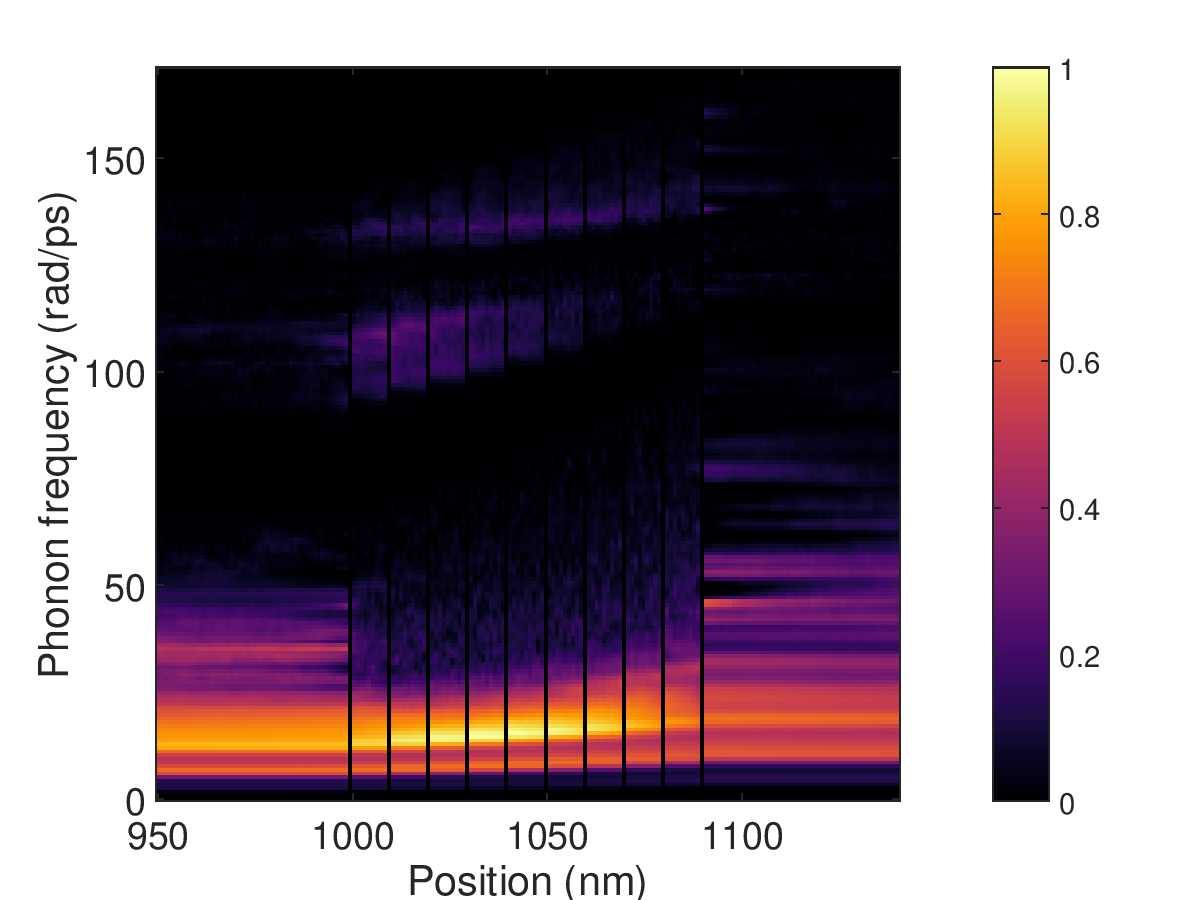}
\par\end{centering}
\caption{Energy-resolved spectral heat flux in a GaN/AlN (001) step graded interface of 90 nm, as computed from Monte Carlo simulations. Upper panel: full structure including the buffer layers; lower panel: zoom close to the interface. The spectral heat flux has been normalized to its maximum value, \SI{3.4}{\micro J / (m^2.rad)}.\label{fig:Spectral_flux}}

\end{figure}

In the limit of an infinitely long interface, the resistance should follow an asymptotic trend equal to the total length divided by the average of the bulk conductivities of the different alloys, here \SI{25.9}{W/(m.K)} compared to the bulk conductivity of the effective alloy of \SI{22.8}{W/(m.K)}. In the limit of an infinitely thin interface, the conductivity becomes limited by the transmission probability and the resistance of an abrupt interface is computed to be \SI{2}{m^2.K / GW}. In contrast, a step graded interface of 9 nm already has a resistance of \SI{3.4}{m^2.K / GW}, showing that the alloy scattering will start to dominate for longer interfaces. Interestingly, this latter value is very much in line with the resistance of the effective alloy (see middle panel of Figure~\ref{fig:Resistance}). Indeed, at these small length scales the resistance originating from mismatch scattering is partially compensated by a lowering of the resistance caused by alloy scattering, due to the simultaneous filtering of phonon modes at the successive interfaces.

This effect is illustrated by Figure~\ref{fig:Spectral_flux}, which displays the energy-resolved spectral heat flux of a structure including a GaN/AlN (001) step graded interface of \SI{90}{nm}. The successive interfaces are very transparent to the main conduction channels at low energies, while the alloy scattering strongly limits the mean free path of the optical modes around \SI{50}{rad/ps} compared to the pure materials. The high-frequency features above \SI{100}{rad/ps} are due to the redistribution of phonon population, since their mean-free path is barely impacted by alloy scattering compared to the aforementioned modes around 50 rad/ps. This is due to the different character of those modes: the alloy scattering rates computed within the Tamura formula\cite{Tamura_1984} are linked to the density of states of Ga/Al character while the highest modes are mostly of N character, as pointed out in the case of InN by Polanco and Lindsay \cite{Polanco_Lindsay_InN}. In the framework of the Monte Carlo solver within the Random Time Approximation, the flux is thus redistributed proportionally more in those modes in the alloys than in the pure compounds.

\section{Optimizing thermal conductance in real devices}

We now discuss the potential optimization of such interfaces. On the one hand, making the graded interface thicker lowers the density of threading dislocations. This improves overall performance in a device-dependent fashion that can be complex to quantify. On the other hand, thicker interfaces can also reduce performance because of device heating resulting from an increased thermal resistance. The balance between these counteracting effects will lead to an optimal grading thickness, which should be optimized on a case-by-case basis for each device. Some current devices are grown on silicon substrate that can be \SI{1000}{\micro m}, which adds a thermal resistance of $\sim$\SI{7700}{m^2.K/GW}. In such cases the device thermal resistance is dominated by the substrate, making it advantageous to have thick graded regions, since the gain in electrical performance by reducing dislocations is greater than the thermal penalty they contribute.

The picture is very different, however, if a good thermally conducting substrate is employed. For example, diamond's thermal conductivity is 17 times higher than silicon's, translating into a $\sim$\SI{450}{m^2.K/GW} thermal resistance for \SI{1000}{\micro m} thickness. As Figure~\ref{fig:Resistance} shows, in this case the resistance contributed by the graded interfaces in the device would already be a non-negligible part of the total. A difficulty in making good conducting substrates is related to the thermal resistance of the bonding between the substrate and the device. However, very low thermal resistances have recently been achieved for diamond/GaN and SiC/GaN, of $36$ and $4.1$ \SI{}{m^2.K/GW} respectively \cite{cho_low_2012,cho_improved_2013,radway_near_2017}. On such new-generation substrates it can pay off to reduce the thickness of the graded regions to some extent, at the expense of an increased dislocation density, to decrease overheating and enhance efficiency and device lifetime.

An increased dislocation density can also lead to increased thermal resistance. However, \textit{ab initio} calculations and analysis of experimental data indicate that, for dislocation densities under \SI{1e10}{cm^{-2}} the thermal conductivity reduction in GaN due to dislocations is minimal, and has been overshadowed by thin film effects in previous experiments \cite{wang_phonon_2018}. The threading dislocation density in GaN directly grown on sapphire, SiC, or Si (111) is typically between \SIrange[range-phrase=--, range-units=single]{1e8}{1e10}{cm^{-2}} \cite{ashby_low-dislocation-density_2000}.

The mean-time-to-failure (MTF) of a device is affected by different mechanisms, and its temperature dependence depends on the particular device in hand \cite{burnham_reliability_2017,cheney_degradation_2012,sun_progressive_2015,coffie_temperature_2007}. Common temperature dependences of GaN device MTFs follow exponential behaviors, and may decrease by one order of magnitude for every $\sim$50 \degree{}C increase in the temperature of the active region. The operation temperature of High Electron Mobility Transistors can reach well above one hundred degrees Celsius. In the well-bonded diamond substrate case above, the over 20\% thermal resistance contributed by a \SI{1}{\micro m}-thick graded interface may thus lead to overheating by \SI{20}{\celsius} and shorten the device's MTF by half.

\section{Conclusion}

In conclusion, using an \textit{ab initio} Green's function approach we have computed the lattice thermal resistances of a series of GaN/AlN (001) step graded interfaces of different lengths. We find that mismatch scattering is very weak in the main conduction channels, such that alloy scattering becomes dominant in the regime relevant for real interfaces (\SI{10}{nm} to \SI{1}{\micro m}). The computed resistance of the step graded interface is well approximated by that of a model thin-film of an effective alloy, and it strongly deviates from the classical Fourier law predictions. The resistance of \SI{1}{\micro m} and \SI{100}{nm} thick interfaces exceeds conventional
estimates by twofold and fourfold respectively. In the case of well-sinked devices, graded interfaces add a non-negligible contribution to overheating, and it can be advisable to shorten the thickness of the step graded interfaces at the expense of a larger density of threading dislocations. The  thermal resistances calculated in this article represent a valuable information for the optimization of GaN-based power electronic devices.

\begin{acknowledgments}
We acknowledge support from the European Union's Horizon 2020 Research and Innovation Programme, grant number 645776 (ALMA). We thank C. Giesen for sharing thickness data from real substrates.
\end{acknowledgments}

\bibliographystyle{apsrev4-1}
\bibliography{bib}

\end{document}